\documentclass[prc,showpacs,superscriptaddress,twocolumn]{revtex4}

\usepackage{graphicx}
\usepackage{amsmath}
\usepackage{array}[=2016-10-06]

\begin{document}
	
	\title{A deep learning approach for predicting multiple observables in Au+Au collisions at RHIC}
	
	\author{Jun-Qi Tao}
	\email[]{taojunqi@mails.ccnu.edu.cn}
	\affiliation{Key Laboratory of Quark \& Lepton Physics (MOE) and Institute of Particle Physics, Central China Normal University, Wuhan 430079, China}
	\affiliation{Helmholtz Research Academy Hesse for FAIR (HFHF), GSI Helmholtz Center for Heavy Ion Research, Frankfurt am Main 60438, Germany}
	\affiliation{Institute for Theoretical Physics, Johann Wolfgang Goethe University, Frankfurt am Main 60438, Germany}
	
	\author{Xiang Fan}
	\email[]{xfan@mails.ccnu.edu.cn}
	\affiliation{Key Laboratory of Quark \& Lepton Physics (MOE) and Institute of Particle Physics, Central China Normal University, Wuhan 430079, China}
	
	\author{Yang Liu}
	\email[]{liuyang01@mails.ccnu.edu.cn}
	\affiliation{Key Laboratory of Quark \& Lepton Physics (MOE) and Institute of Particle Physics, Central China Normal University, Wuhan 430079, China}
	
	\author{Yu Sha}
	\email[]{yusha@cuhk.edu.cn}
	\affiliation{School of Science and Engineering, The Chinese University of Hong Kong, Shenzhen (CUHK-Shenzhen), Guangdong, 518172, China}
	
	\author{Kai Zhou}
	\email[]{zhoukai@cuhk.edu.cn}
	\affiliation{School of Science and Engineering, The Chinese University of Hong Kong, Shenzhen (CUHK-Shenzhen), Guangdong, 518172, China}
	\affiliation{School of Artificial Intelligence, The Chinese University of Hong Kong, Shenzhen (CUHK-Shenzhen), Guangdong, 518172, China}
	
	\author{Hua Zheng}
	\email[]{zhengh@snnu.edu.cn}
	\affiliation{School of Physics and Information Technology, Shaanxi Normal University, Xi’an 710119, China}
	
	\author{Ben-Wei Zhang}
	\email[]{bwzhang@mail.ccnu.edu.cn}
	\affiliation{Key Laboratory of Quark \& Lepton Physics (MOE) and Institute of Particle Physics, Central China Normal University, Wuhan 430079, China}

	\date{\today}
	
	\begin{abstract}
		
		We develop a neural network model, based on the processes of high-energy heavy-ion collisions, to study and predict several experimental observables in Au+Au collisions. We present a data-driven deep learning framework for predicting multiple bulk observables in Au+Au collisions at RHIC energies. A single neural network is trained exclusively on experimental measurements of charged-particle pseudorapidity density distributions, transverse-momentum spectra and elliptic flow coefficients over a broad range of collision energies and centralities. The network architecture is inspired by the stages of a heavy-ion collision, from the quark-gluon plasma to chemical and kinetic freeze-out, and employs locally connected hidden layers and a structured input design that encodes basic geometric and kinematic features of the system. We demonstrate that these physics-motivated choices significantly improve test performance compared to purely fully connected baselines. The trained model is then used to predict the above observables at collision energies not yet explored experimentally at RHIC, and the results are validated using the energy dependence of the total charged-particle multiplicity per participant pair as well as comparisons to a CLVisc hydrodynamic calculation with TRENTo initial conditions. Our findings indicate that such physics-guided neural networks can serve as efficient surrogates to fill critical data gaps at RHIC and to support further phenomenological studies of QGP properties.
		
	\end{abstract}
	
	
	\maketitle
	
	\section{Introduction}\label{Sec1}
	
	\begin{figure*}
		\includegraphics[scale=0.35]{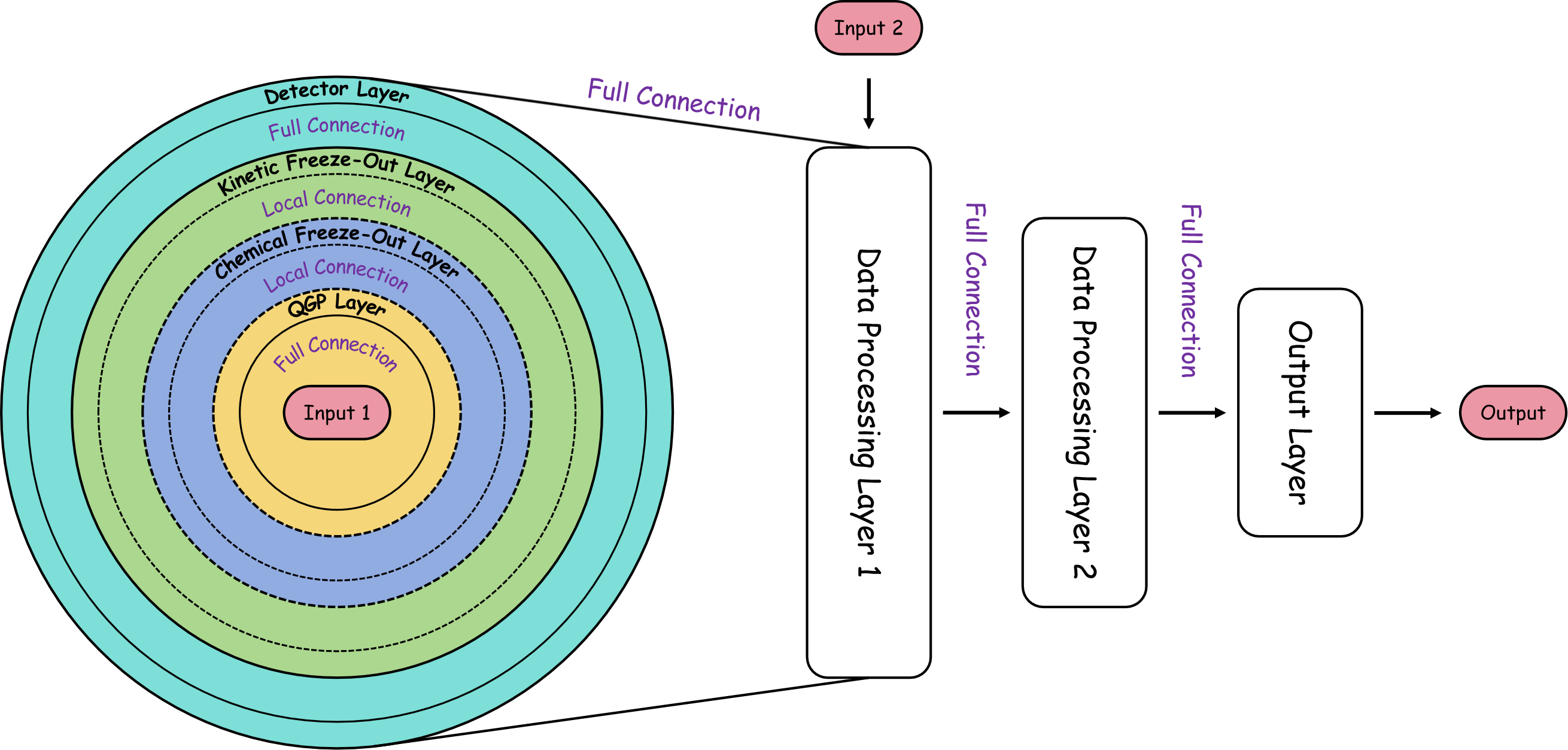}
		\caption{\label{model}The scheme of the neural network model.}
	\end{figure*}
	
	High-energy heavy-ion collision experiments provide a unique way to explore the properties of matter at extreme conditions. In the experiments, the charged particle pseudorapidity density distribution ($dN_{ch}/d\eta$ vs. $\eta$),  transverse momentum dependence of charged particle elliptic flow ($v_2$ vs. $p_T$) and charged particle transverse momentum distribution ($dN_{ch}/2\pi p_T dp_T d\eta$ vs. $p_T$) are important observables for studying the collision processes. They are used to constrain the models developed for studying heavy-ion collisions, such as HIJING \cite{Wang1991}, EPOS-LHC \cite{Pierog2015}, AMPT \cite{Lin2005,Wang2019}, PACIAE \cite{Lei2025, Xie2025}, a multi-source thermal model \cite{Li2014, Sun2013}, hydrodynamic model \cite{Pang2018,Wu2022,Jiang2023,Jiang2023a} et al., and also improve our understanding of the Quark-Gluon Plasma (QGP) characteristics, the properties of the final particles produced in the collisions, the particle production mechanisms and so on.
	
	Due to advanced technology, the development of GPUs and the abundance of big data, machine learning has been developed considerably in the last few years and has been increasingly applied to scientific researches including high-energy nuclear physics \cite{eZhou2024,He2023,Oliveira2017,Steinheimer2019,Huang2022,Mallick2023,OmanaKuttan2020,Pang2019,Mengel2023,Guo2023,Sun2024, Paganini2018, Sun2025, Gao2021, Cao2023, Mumpower2022,eDu2019,eKuttan2021,eZhao2022,eOmana2023}. In Ref. \cite{Oliveira2017}, a novel Generative Adversarial Networks architecture (GAN) is applied to the production of jet images. In Ref. \cite{Steinheimer2019}, a state-of-the-art machine learning method is adopted to study the coordinate and momentum space configurations of the net baryon number in heavy ion collisions that undergo spinodal decomposition, due to a first-order phase transition. In Ref. \cite{Huang2022}, a point cloud network with dynamical edge convolution is employed to identify events with critical fluctuations through supervised learning, and pick out a large fraction of signal particles used for decision-making in each single event. In Ref. \cite{Mallick2023}, a deep learning feed-forward network is used for estimating elliptic flow coefficients in heavy-ion collisions. In Ref. \cite{OmanaKuttan2020}, a deep learning-based event characterization method is proposed for fast, online event-by-event impact parameter determination at the CBM experiment. These studies demonstrate that machine learning has become a powerful tool for studies in high-energy nuclear physics.
	
	In experiments, abundant experimental data in Au+Au collisions have been measured by collaborations at the RHIC, e.g., the charged particle pseudorapidity density distributions, the particle transverse momentum distributions and the transverse momentum dependence of the elliptic flow of particles \cite{Back2003,Back2006,Alver2011,Adamczyk2012,Adler2002,Adare2010,Abelev2010,Adam2020,Abelev2007,Adamczyk2018,Adams2003,Adler2002a}. However, the collision energy points are still very limited due to practical reasons, such as physics goals, cost and time. In this paper, a neural network model is built based on the high-energy heavy-ion collision experiment processes to study and predict several experimental observables for which the experimental data are not available. This neural network model is trained by using the pseudorapidity density distributions, elliptic flow and transverse momentum distributions of charged particles in Au+Au collisions at the RHIC. The model, which incorporates physical insights into its design, is also compared with alternative approaches and demonstrates better performance. Then the trained neural network is used to make predictions about these observables for unexplored energies. The collision energy $\sqrt{s_{NN}}$ dependence of the total charged particle multiplicity per participant pair is also used to assess the accuracy of the predictions.
	
	The paper is organized as follows. In section \ref{Sec2}, we introduce the neural network model. In Section \ref{Sec3}, we present the comparison of the results from our model and alternative approaches. The training, test results and predictions of pseudorapidity density distribution, elliptic flow and transverse momentum distribution of charged particles given by the neural network model are shown. The verification results of predictions using the total charged particle multiplicity per participant pair are presented. The dependencies of the predicted total charged particle multiplicity on the number of participants are also discussed. In section \ref{Sec4}, a brief conclusion is drawn.

	\section{Model framework}\label{Sec2}

	High-energy heavy-ion collision experiments can be generalized into the following stages. First, two heavy-ions colliding at high-energy undergo a preequilibrium stage and produce the high-temperature and high-density matter, QGP. Second, after evolution, the QGP begins hadronization and reaches chemical freeze-out, i.e. the particle composition is essentially fixed. Third, the particles that have reached chemical freeze-out gradually stop interacting and reach kinetic freeze-out. Fourth, the particles that have reached kinetic freeze-out continue on and eventually reach the detector. Then the data are analyzed and presented as the final experimental results, such as the pseudorapidity density distributions, elliptic flow and transverse momentum spectra of particles.
	
	A neural network model is built to mimic the high-energy heavy-ion collision experiment procedures described above (see Fig. \ref{model}). The model architecture includes six hidden layers: the QGP layer, chemical freeze-out layer, kinetic freeze-out layer, detector layer, and two data processing layers, as well as two input layers and one output layer, corresponding to the different stages of the experiment. Considering the size of the training set and the performance of the neural network, instead of making the neural network directly output the dependence of the observables on the final particle kinematics parameters after inputting the collision system initial information (i.e. input 1, such as (196.97, 196.97, 19.6, 0, 10), where 196.97 and 196.97 are the relative atomic masses of the two colliding particles, 19.6 is the collision energy in units of GeV, 0 and 10 are the collision centrality range limits) into the neural network, we input the final particle kinematics parameters and corresponding one-hot encoding (i.e. input 2, such as pseudorapidity ($\eta, 1, 0, 0$))into the neural network as well at data processing layer 1, and make the neural network output observables (i.e. output, such as pseudorapidity density distributions ($dN/d\eta$)) corresponding to the final particle kinematics parameters. We assume the evolution from the QGP stage to the kinetic freeze-out stage as a uniformly expanding fireball. Therefore, we distribute the neurons in the QGP, chemical freeze-out and kinetic freeze-out layers uniformly on the spheres with increasing radius. The method of uniform distribution is the Fibonacci grid,
	\begin{equation}
		\begin{aligned}
			z_n &= \frac{R(2n-1)}{N-1},\\
			x_n &= \sqrt{R^2-z_n^2}\cos(2\pi n\phi),\\
			y_n &= \sqrt{R^2-z_n^2}\sin(2\pi n\phi),
		\end{aligned}
	\end{equation}
	where $z_n, x_n, y_n$ are the coordinates of the $n$th neuron on the sphere centered at the coordinate origin, $R$ is the radius of the sphere, $N$ is the total number of neurons in the layer, $\phi$ is the golden ratio. With this distribution, we can determine the distance between neurons and implement the local connection between the QGP, chemical freeze-out and kinetic freeze-out layers. Each neuron in the chemical freeze-out layer is connected only to its seven nearest neighbors in the QGP layer, and the same principle is applied to the connection between the chemical freeze-out layer and kinetic freeze-out layer. This induces a structured, sparse connectivity pattern that regularizes the network by suppressing spurious long-range couplings while preserving local correlations, playing a similar role to Dropout but in a more physically interpretable way. For the connection between other layers, the full connection is used.
	
	The neural network model is implemented with PyTorch \cite{Paszke2019}. The model consists of layers with neuron counts of 200, 300, 400, 400, 404, 304, and 204 from the QGP layer to the output layer, respectively. Leaky ReLU \cite{Maas2013} is used as the activation function for the hidden layers, while no activation function is applied to the output layer. The model is optimized using the ADAM optimizer \cite{Kingma2017}, and Huber loss is used as the loss function,
	\begin{equation}
		\begin{aligned}
			L(y-\hat{y}) =
			\begin{cases}
				\frac{1}{2} (y-\hat{y})^2, & \text{if } |y-\hat{y}| \leq 1 \\
				 |y-\hat{y}| - \frac{1}{2}, & \text{if } |y-\hat{y}| > 1
			\end{cases}
		\end{aligned}
	\end{equation}
	where $y$ is the experimental data and $\hat{y}$ is the prediction given by the neural network. The training set include charged particle pseudorapidity density distributions in Au+Au collisions at $\sqrt{s_{NN}}=19.6, 62.4, 200$ GeV \cite{Back2003,Back2006}, transverse momentum dependence of charged particle elliptic flow in Au+Au collisions at $\sqrt{s_{NN}}=7.7, 9.2, 11.5, 14.5, 27, 62.4, 130, 200$ GeV \cite{Adamczyk2012,Adler2002,Adare2010,Abelev2010, Adam2020, Abelev2007}, and transverse momentum distribution of charged particles produced in Au+Au collisions at $\sqrt{s_{NN}}=7.7, 11.5, 14.5, 27, 39, 62.4, 200$ GeV \cite{Adamczyk2018,Adams2003}. The test set include charged particle pseudorapidity density distributions in Au+Au collisions at $\sqrt{s_{NN}}=130$ GeV \cite{Back2003}, transverse momentum dependence of charged particle elliptic flow in Au+Au collisions at $19.6, 39$ GeV \cite{Adamczyk2012}, and transverse momentum distribution of charged particles produced in Au+Au collisions at $19.6, 130$ GeV \cite{Adamczyk2018,Adler2002a}. See Table. \ref{set} for clarity. All three kinds of experimental data are simultaneously used to train and test the neural network model.
	
	\begin{table*}[!htb]
		\renewcommand{\arraystretch}{1.5}
		\centering
		\caption{The experimental data included in the training and test set.}
		\begin{tabular}{>{\centering\arraybackslash}p{5.5cm}
				>{\centering\arraybackslash}p{5.5cm}
				>{\centering\arraybackslash}p{5.5cm}}
			\hline\hline
			\textbf{Observables (Charged)} &
			\textbf{Collision energy in the training set (GeV)} &
			\textbf{Collision energy in the test set (GeV)} \\
			\hline
			pseudorapidity density distributions & $19.6, 62.4, 200$ & $130$ \\
			Elliptic flow                        & $7.7, 9.2, 11.5, 14.5, 27, 62.4, 130, 200$ & $19.6, 39$ \\
			Transverse momentum distributions    & $7.7, 11.5, 14.5, 27, 39, 62.4, 200$        & $19.6, 130$ \\
			\hline\hline
		\end{tabular}
		\label{set}
	\end{table*}

	\section{Results}\label{Sec3}
	
	\begin{figure}
		\includegraphics[scale=0.35]{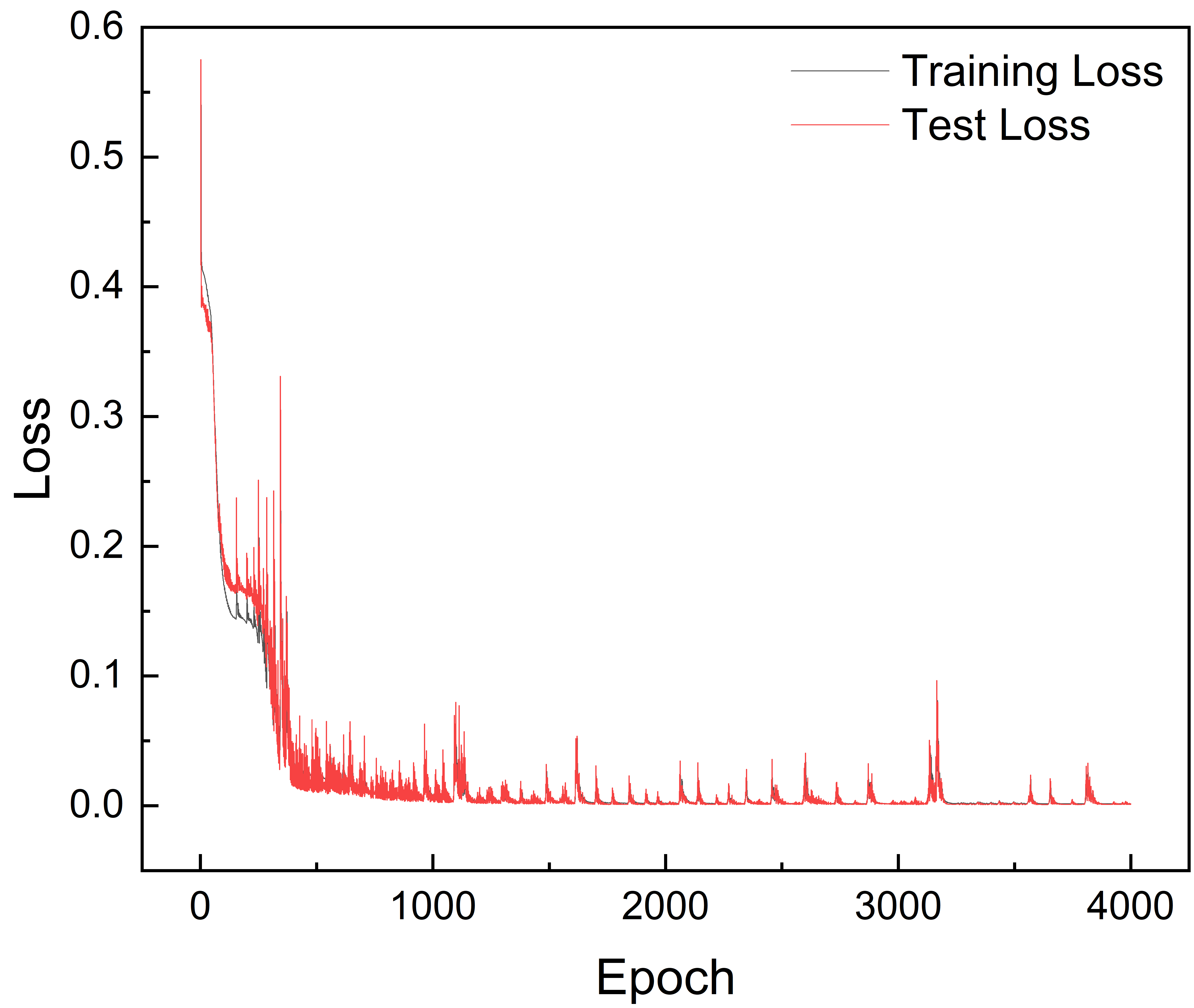}
		\caption{\label{loss}The training and test loss of the neural network model as a function of the training epochs.}
	\end{figure}
	
	\begin{table*}[!htb]
		\renewcommand{\arraystretch}{1.5} 
		\centering
		\caption{The best training and test loss for different neural network models}
		\begin{tabular}{>{\centering\arraybackslash}p{10cm}
				>{\centering\arraybackslash}p{3cm}
				>{\centering\arraybackslash}p{3cm}}
			\hline\hline
			\textbf{Neural network models} &
			\textbf{Training loss} &
			\textbf{Test loss} \\
			\hline
			Model used & $1.41\times10^{-3}$ & $0.71\times10^{-3}$ \\
			Model without local connection & $5.72\times10^{-3}$ & $3.77\times10^{-3}$ \\
			Model without splitting the input & $3.18\times10^{-3}$ & $1.61\times10^{-3}$ \\
			Model without local connection and splitting the input & $17.65\times10^{-3}$ & $4.88\times10^{-3}$ \\
			Model replacing the locally connection with Dropout layers & $3.35\times10^{-3}$ & $1.07\times10^{-3}$ \\
			\hline\hline
		\end{tabular}
		\label{d_loss}
	\end{table*}
	
	\begin{figure*}
		\includegraphics[scale=0.25]{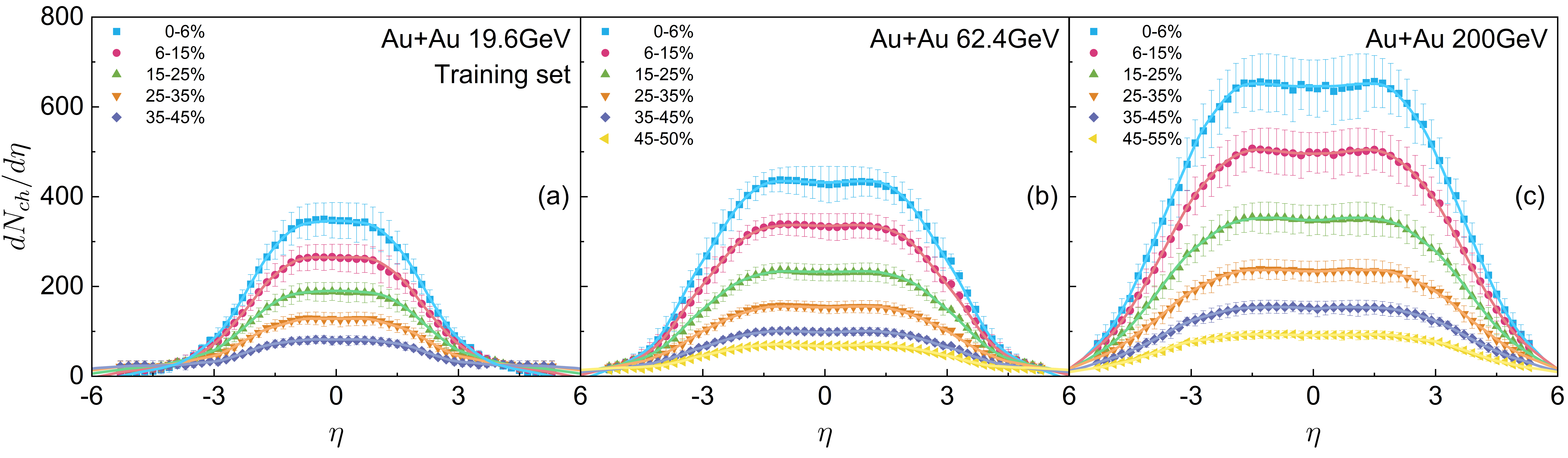}
		\caption{\label{eta_train}The charged particle pseudorapidity density distributions produced in Au + Au collisions at $\sqrt{s_{NN}}=19.6, 62.4$ and $200$ GeV for different centralities. The symbols are experimental data taken from Refs. \cite{Back2003,Back2006}. The curves are the training results of the neural network.}
	\end{figure*}
	
	\begin{figure}
		\includegraphics[scale=0.35]{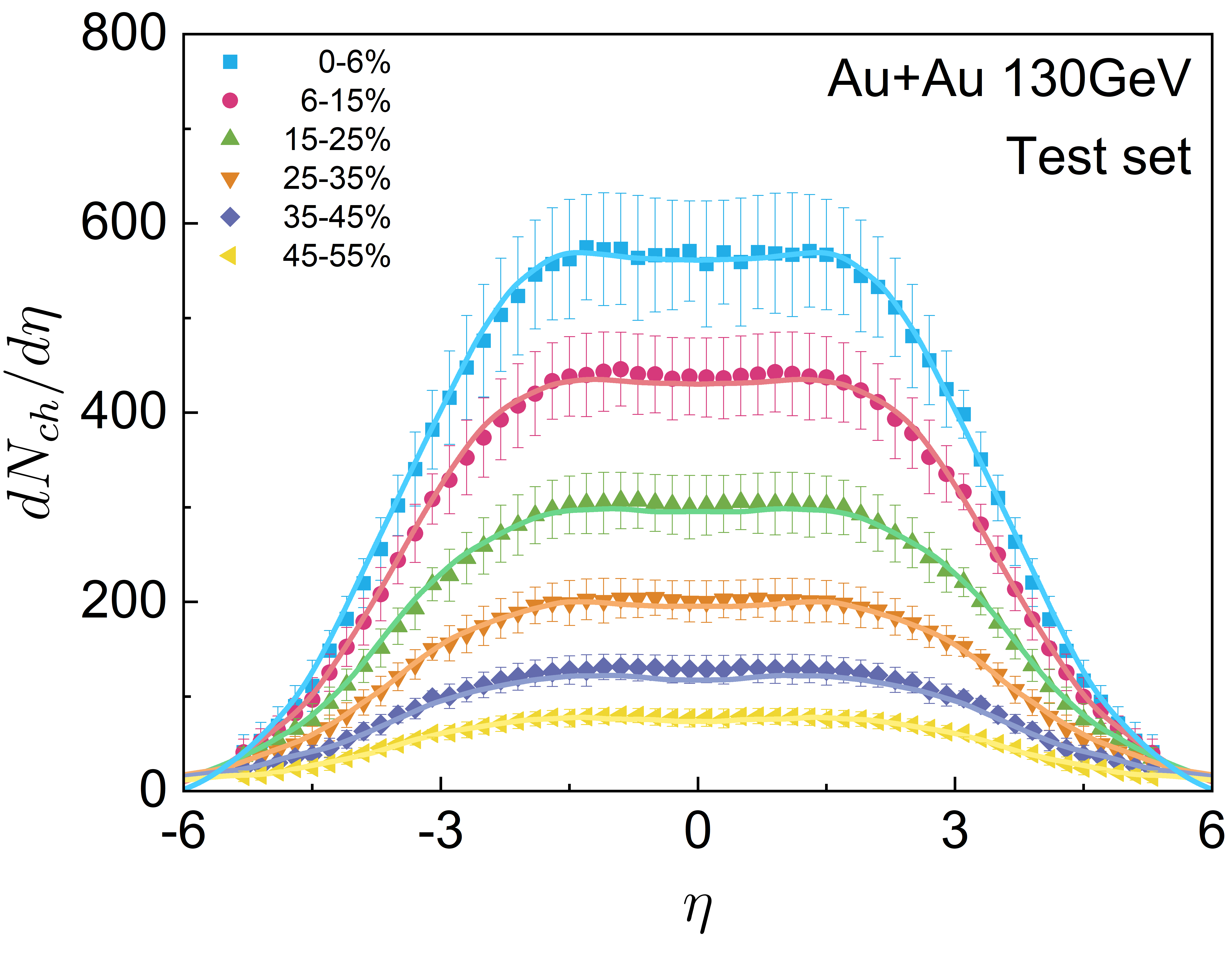}
		\caption{\label{eta_test}The charged particle pseudorapidity density distributions produced in Au + Au collisions at $\sqrt{s_{NN}}=130$ GeV for different centralities. The symbols are experimental data taken from Ref. \cite{Back2003}. The curves are the test results of the neural network.}
	\end{figure}
	
	\begin{figure*}
		\includegraphics[scale=0.25]{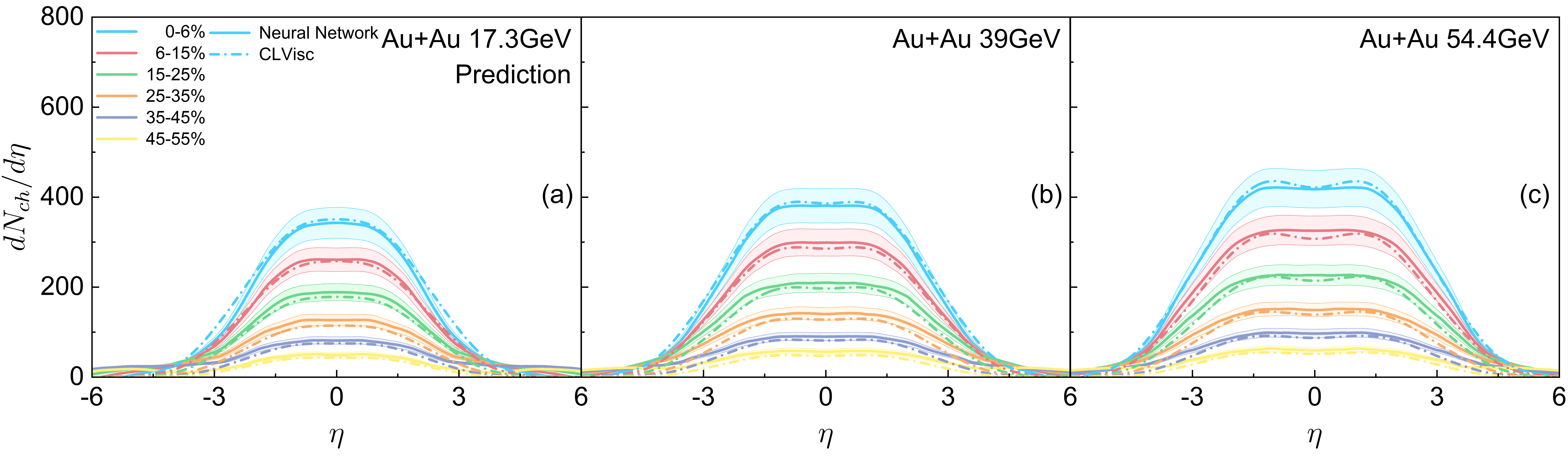}
		\caption{\label{eta_predicition}The charged particle pseudorapidity density distributions produced in Au + Au collisions at $\sqrt{s_{NN}}=17.3, 39$ and $54.4$ GeV for different centralities. The curves are the predictions of the neural network. The dash-dotted lines are the results of CLVisc. The bands are the error bars.}
	\end{figure*}
	
	\begin{figure}
		\includegraphics[scale=0.7]{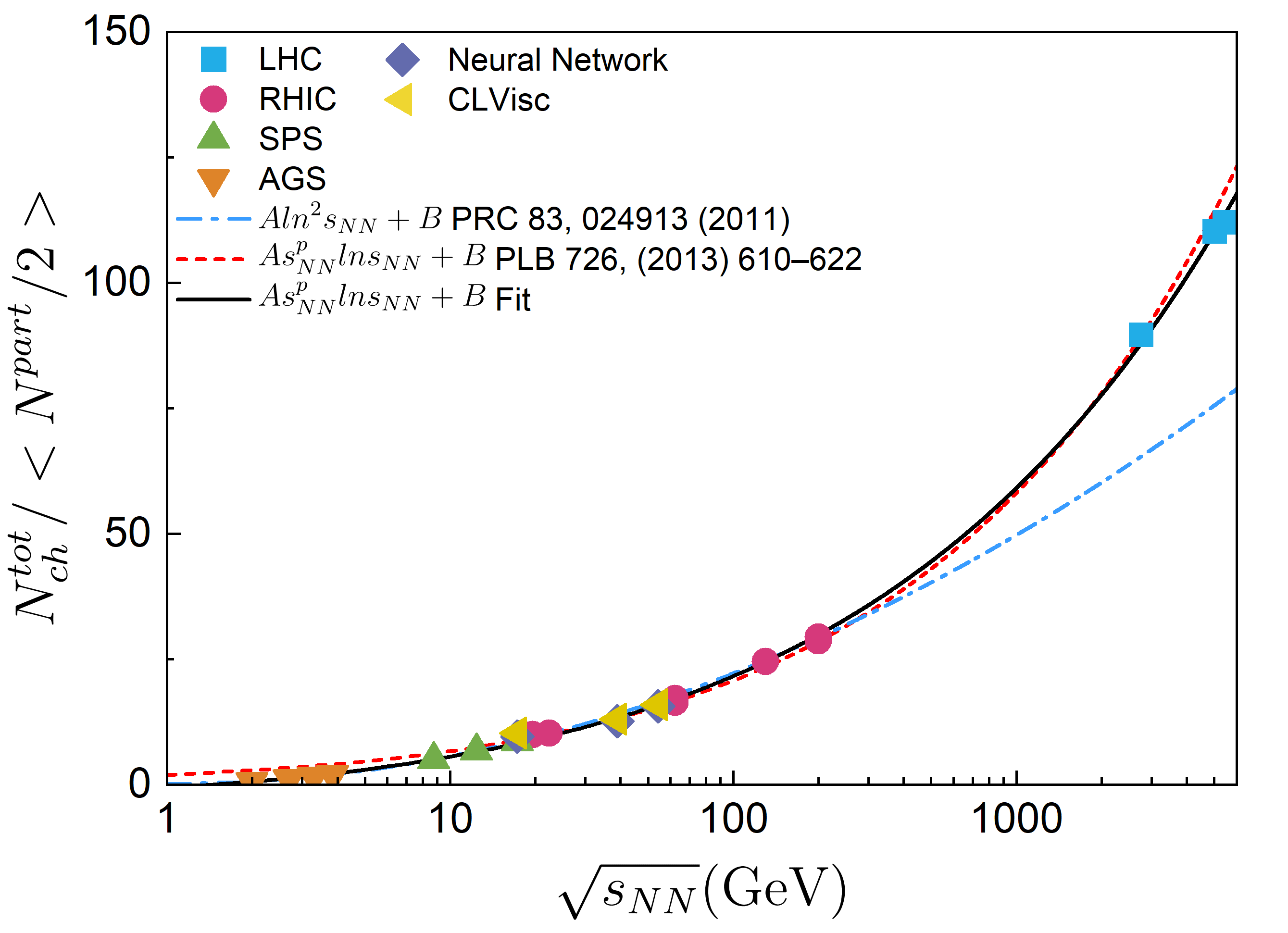}
		\caption{\label{totalN_Npart}The total charged particle multiplicity per participant pair in Au+Au collisions versus $\sqrt{s_{NN}}$ for the most central centrality. The symbols representing experimental data from AGS (0-5\% Au+Au) \cite{Klay2003,Ahle1998}, SPS (0-5\% Pb+Pb) \cite{Afanasiev2002,Abreu2002}, RHIC (0-6\%, Au+Au and Cu+Cu) \cite{Back2003,Back2006,Alver2011}, LHC (0-5\%, Pb+Pb and Xe+Xe) \cite{Abbas2013,Adam2017,Acharya2019}. The diamonds and left-facing triangles are the results of the neural network and CLVisc, respectively. The dashed and dash-dotted lines are the results from fit to lower energy data \cite{Alver2011,Abbas2013}. The full line is the result of fit over all experimental data, excluding the results of the neural network and CLVisc.}
	\end{figure}
	
	\begin{figure}
		\includegraphics[scale=0.6]{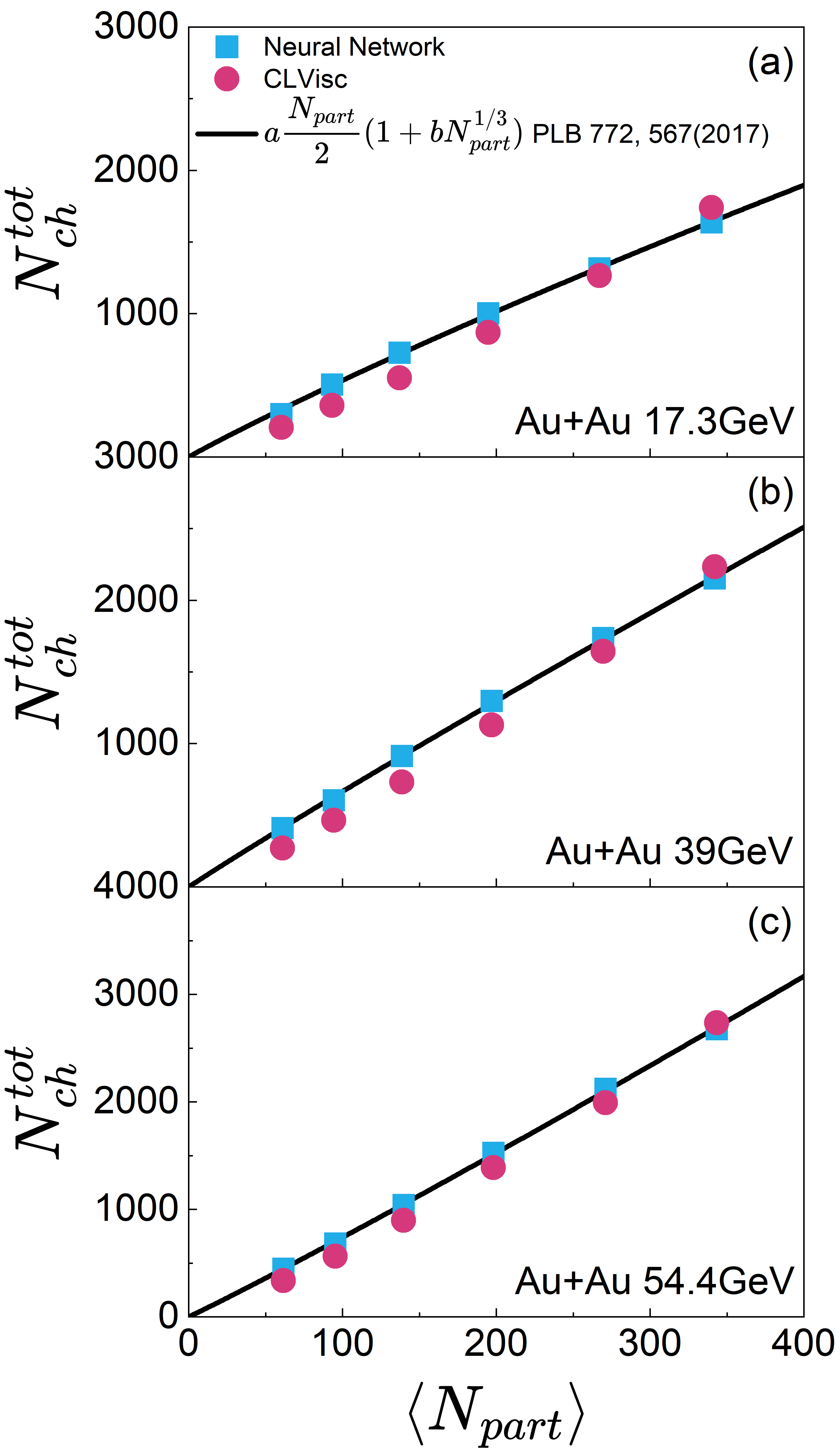}
		\caption{\label{totalN_vs_Npart}The total charged particle multiplicity produced in Au + Au collisions at $\sqrt{s_{NN}}=17.3, 39$ and $54.4$ GeV versus the number of participants. The squares and circles are the results of the neural network and CLVisc. The lines are the fitting results for the predictions of the neural network. The fitting function is from Ref. \cite{Adam2017}.}
	\end{figure}
	
	\begin{figure*}
		\includegraphics[scale=0.2]{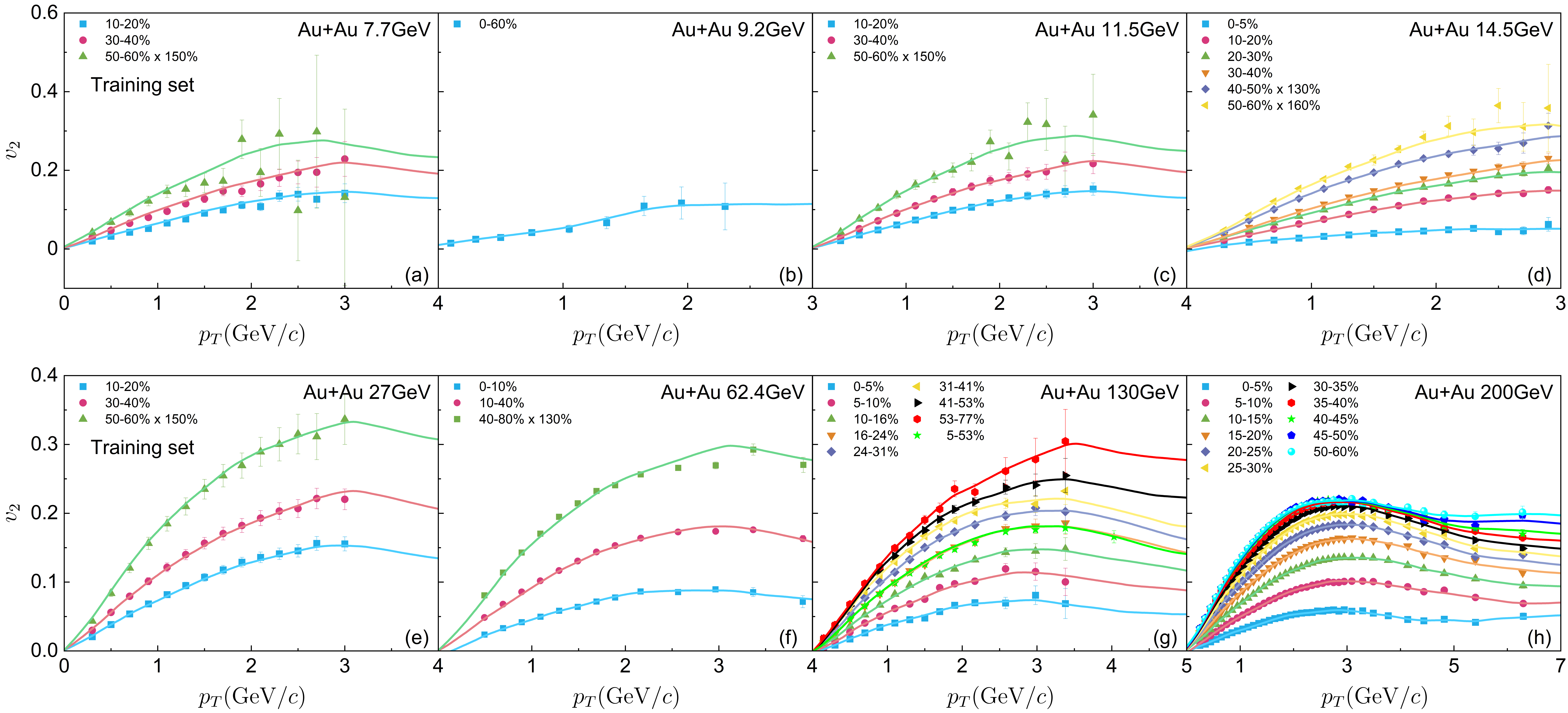}
		\caption{\label{v2_train}The transverse momentum dependence of charged particle elliptic flow in Au+Au collisions at $\sqrt{s_{NN}}=7.7, 9.2, 11.5, 14.5, 27, 39, 62.4, 130$ and $200$ GeV for different centralities. The symbols are experimental data taken from Refs. \cite{Adamczyk2012,Adler2002,Adare2010,Abelev2010, Adam2020, Abelev2007}. The curves are the training results of the neural network.}
	\end{figure*}
	
	\begin{figure*}
		\includegraphics[scale=0.3]{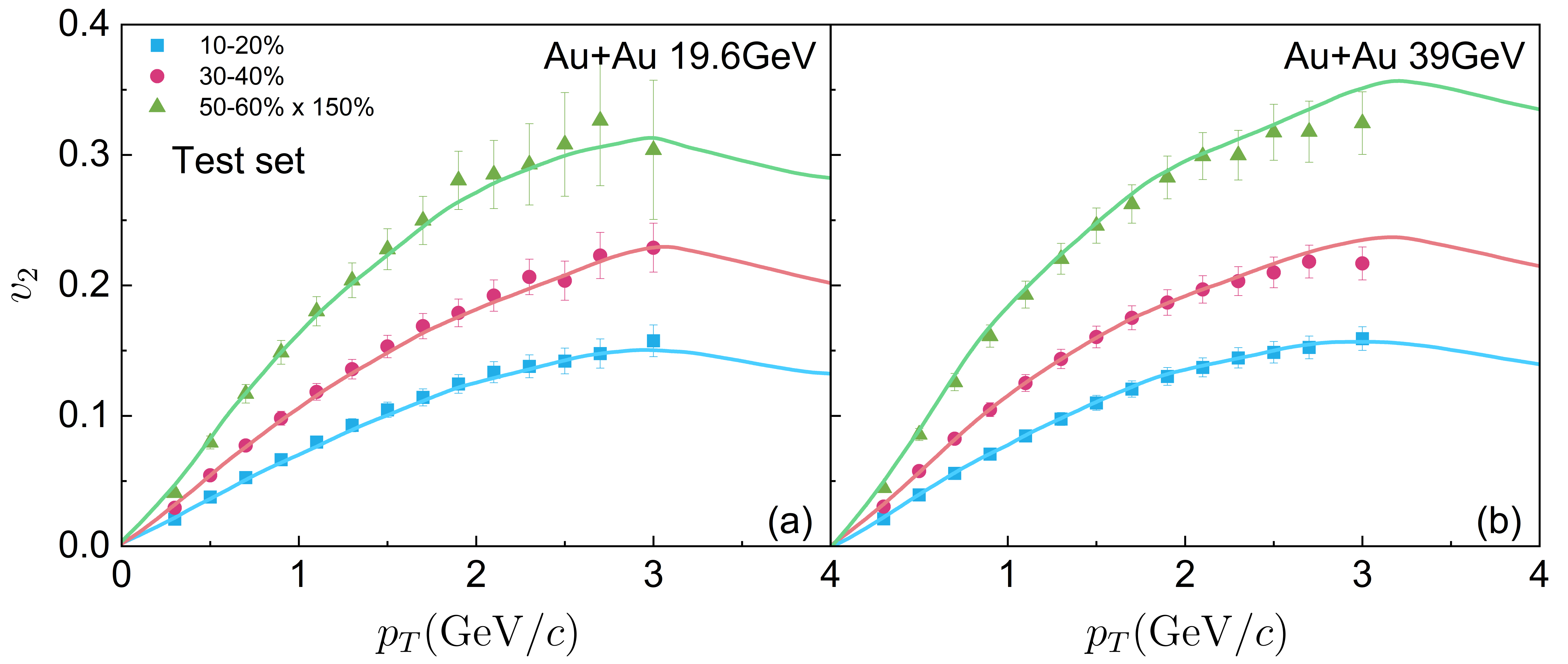}
		\caption{\label{v2_test}The transverse momentum dependence of charged particle elliptic flow in Au+Au collisions at $\sqrt{s_{NN}}=19.6$ and $39$ GeV for different centralities. The symbols are experimental data taken from Ref. \cite{Adamczyk2012}. The curves are the test results of the neural network.}
	\end{figure*}
	
	\begin{figure*}
		\includegraphics[scale=0.25]{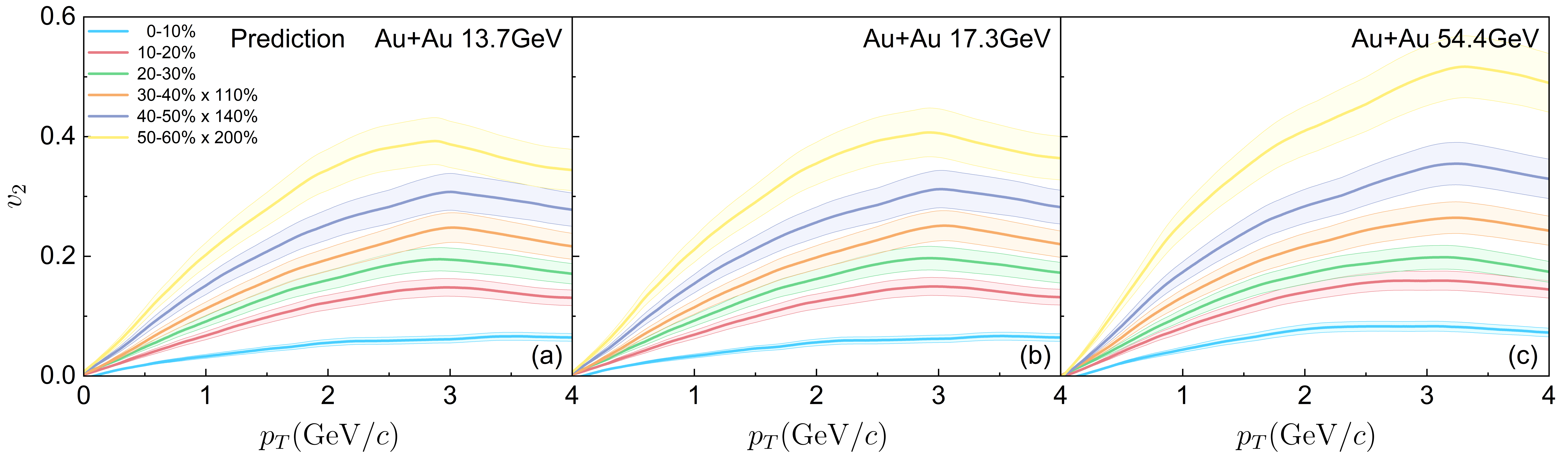}
		\caption{\label{v2_prediction}The transverse momentum dependence of charged particle elliptic flow in Au+Au collisions at $\sqrt{s_{NN}}=13.7, 17.3$ and $54.4$ GeV for different centralities. The curves are the predictions of the neural network. The bands are the error bars.}
	\end{figure*}
	
	\begin{figure*}
		\includegraphics[scale=0.18]{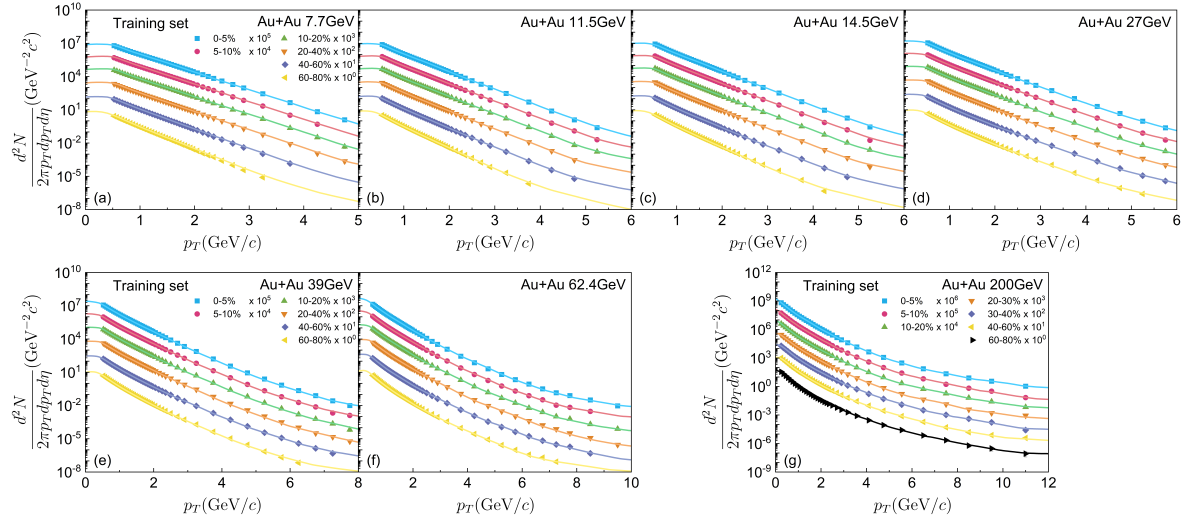}
		\caption{\label{pt_train}The transverse momentum distributions of charged particles produced in Au+Au collisions at $\sqrt{s_{NN}}=7.7, 11.5, 14.5, 27, 39, 62.4$ and $200$ GeV for different centralities. The symbols are experimental data taken from Refs. \cite{Adamczyk2018,Adams2003}. The curves are the training results of the neural network. The bands are the error bars.}
	\end{figure*}
	
	\begin{figure*}
		\includegraphics[scale=0.3]{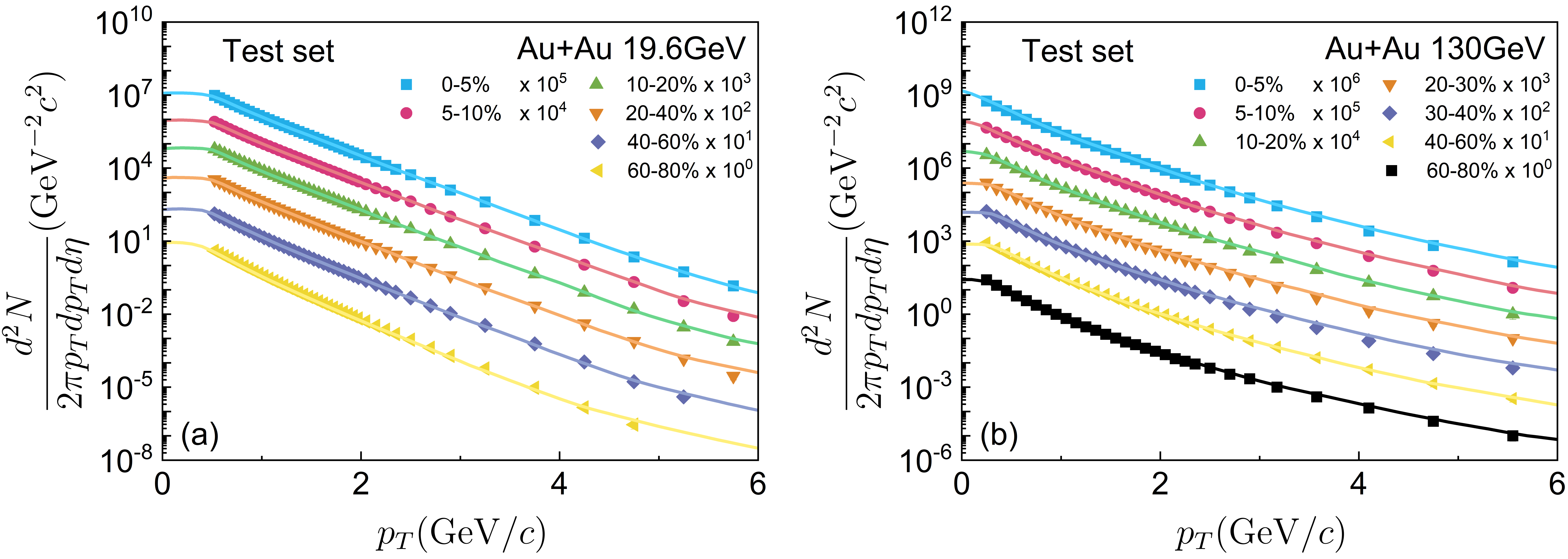}
		\caption{\label{pt_test}The transverse momentum distributions of charged particles produced in Au+Au collisions at $\sqrt{s_{NN}}=19.6$ and $130$ GeV for different centralities. The symbols are experimental data taken from Refs. \cite{Adamczyk2018,Adler2002a}. The curves are the test results of the neural network.}
	\end{figure*}
	
	\begin{figure*}
		\includegraphics[scale=0.25]{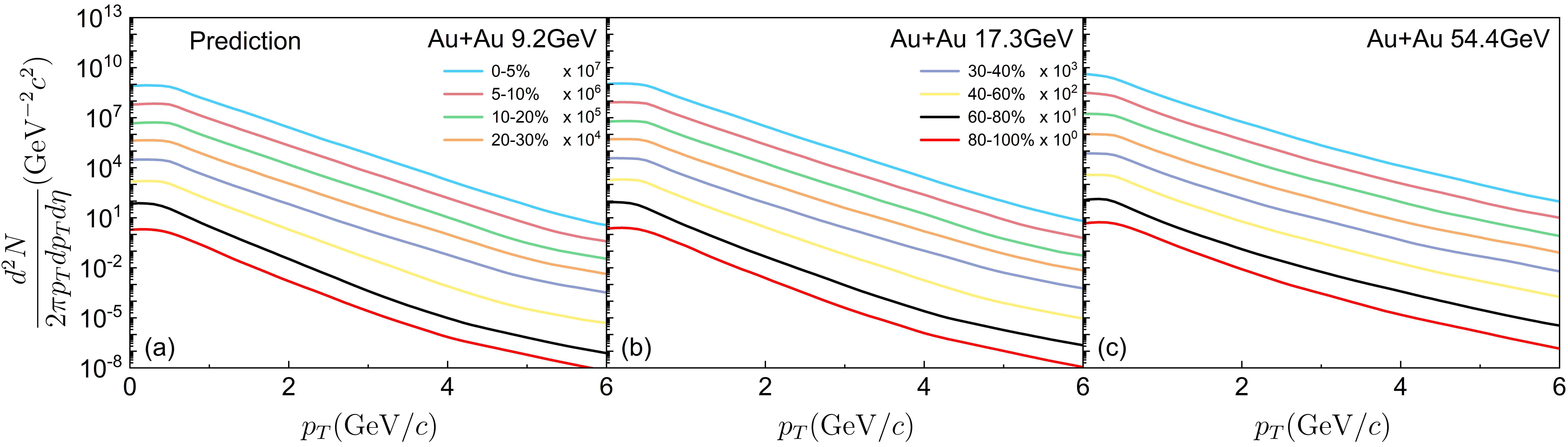}
		\caption{\label{pt_prediction}The transverse momentum distributions of charged particles produced in Au+Au collisions at $\sqrt{s_{NN}}=9.2, 17.3$ and $54.4$ GeV for different centralities. The curves are the predictions of the neural network. The bands are the error bars.}
	\end{figure*}
	
	In recent studies \cite{Mallick2023,OmanaKuttan2020,Pang2019,Mengel2023,Guo2023,Sun2024}, some neural networks have been trained using a combination of experimental data and synthetic data produced by different models, such as hydrodynamic simulations or transport models. This combined approach provides a more diverse set of training data and enhances the generalizability of the models. In contrast, our work exclusively uses experimental data to train the neural network. This approach tests the capability of the neural network model to learn directly from experimental data.
	
	Due to the significant differences in value ranges among the three data categories, a direct combination of the raw data could lead to a biased or ineffective training process. Therefore, we standardized each category independently before combining them into a unified dataset. Additionally, considering that the transverse momentum distribution of charged particles approaches zero in the high $p_T$ region, we applied a logarithmic transformation to the experimental data prior to inclusion in the dataset to ensure the robustness of model performance.
	
	In Fig. \ref{loss}, the training and test loss of the neural network are shown. It is important to note that without employing local connections, or without splitting the input into two parts and feeding them into the neural network from different locations, such satisfactory results cannot be achieved. Even replacing the local connection with Dropout layers of the same drop rate fails to yield comparable performance. This finding aligns well with biological observations where sparse connectivity and specifically structured connections, guided by neuronal morphology, enhance the efficiency and effectiveness of cortical networks \cite{Gregg2021,Daniel2022}. The best training and testing losses for different models are summarized in Tab. \ref{d_loss}, demonstrating that incorporating physical insights plays a crucial role in the design of effective neural network architectures \cite{Mumpower2022}.
	
	In Fig. \ref{eta_train}, the training results of the neural network for the charged particle pseudorapidity density distributions with $\sqrt{s_{NN}}=19.6, 62.4$ and $200$ GeV at different centrality bins in Au+Au collisions are shown. The training results of the neural network are consistent with the experimental data except for the data of the charged particle with $|\eta|>4$ in Au+Au collisions at $\sqrt{s_{NN}}=19.6$ GeV, which coincide with the beam-remnant region where longitudinal fragmentation effects dominate, which are not strongly constrained by our training data. We therefore do not expect the present architecture, which focuses on bulk particle production, to capture this region quantitatively.
	
	In Fig. \ref{eta_test}, the test results of the neural network for the charged particle pseudorapidity density distributions with $\sqrt{s_{NN}}=130$ GeV at different centrality bins in Au+Au collisions are shown. As one can see, the neural network performs well on the test set. This indicates that the neural network has generalization ability and can make valid predictions. It is also worth noting that in the mid-pseudorapidity region, i.e. $|\eta|<1.5$, the results given by the neural network are straight lines, which indicates that the model captures the plateau structure of the charged particle pseudorapidity density distributions at the mid-pseudorapidity.
	
	In Fig. \ref{eta_predicition}, we show the predictions of the neural network and the results given by the CLVisc model for the pseudorapidity density distributions of charged particles with $\sqrt{s_{NN}}=17.3, 39$ and $54.4$ GeV at different centrality bins in Au+Au collisions. Since the error in the experimental data averages around 10\%, an error 10\%  is added to the predictions of the neural network. The CLVisc model parameters are tuned based on the charged particle pseudorapidity density distributions in the most central collision. See Appendix \ref{appendixA} for details. It can be found that the results given by the neural network and CLVisc model at different centrality bins are basically consistent within the errors. This indirectly verifies the reliability of the neural network predictions. We notice that the neural network can give reasonable predictions for collision centrality that do not exist in the training set for low collision energy, which indicates that the neural network has the generalization ability for centrality.
	
	In Fig. \ref{totalN_Npart}, we show the total charged-particle multiplicity per participant pair versus the collision energy for the most central centrality. The full line is the fitting result for all collision energies, excluding neural network and CLVisc model results. The number of participants in Au+Au collisions at $\sqrt{s_{NN}}=17.3, 39$ and $54.4$ GeV is generated by T\raisebox{-0.5ex}{R}ENTo. One can observe that the predictions of the neural network and the results of the CLVisc model are consistent with the fitting result for all collision energies (full line), which demonstrates the reliability of the neural network.
	
	In Fig. \ref{totalN_vs_Npart}, we show the total charged particle multiplicity versus the number of participants. The fitting function from Ref. \cite{Adam2017} can reproduce all the predictions of the neural network well. The number of participants is generated by T\raisebox{-0.5ex}{R}ENTo. We can find that the results given by the CLVisc model are slightly lower than those given by the neural network, except for the most central collisions used to tune the parameters of the CLVisc model. It can also be found that the dependence of the total charged particle multiplicity given by the neural network on the number of participants is similar to that given by the experimental collaborations \cite{Gao2017,Tao2021,Tao2023}, which can to some extent demonstrate the reliability of the neural network indirectly.
	
	In Fig. \ref{v2_train}, the training results of the neural network for the transverse momentum dependence of charged particles elliptic flow with $\sqrt{s_{NN}}=7.7, 9.2, 11.5, 14.5, 27, 39, 62.4, 130$ and $200$ GeV at different centrality bins in Au+Au collisions are shown. It can be seen that the results from the neural network agree well with the experimental data, except for the range with the elliptic flow has large fluctuations, i.e. $p_T\approx2-3$ GeV at $\sqrt{s_{NN}}=7.7, 11.5$ and $14.5$ GeV, since the results from the neural network are smooth similar to the results in Ref. \cite{Wang2022}. In this range, the curves given by the neural network do not pass through these points, but choose an appropriate path, which indicates that the neural network does not overfit the training set of charged particle elliptic flow.
	
	In Fig. \ref{v2_test}, the test results of the neural network for the transverse momentum dependence of charged particle elliptic flow with $\sqrt{s_{NN}}=19.6$ and $39$ GeV at different centrality bins in Au+Au collisions are shown. As one can see, the neural network reproduces the experimental results well. This indicates that the trained neural network achieves reliable performance, enabling it to make accurate predictions. It is also worth noting that the results given by the neural network are convex curves, which indicate that the neural network model effectively captures the characteristic rise, saturation and fall of $v_2$ with increasing $p_T$.
	
	In Fig. \ref{v2_prediction}, we show the predictions of the neural network for the transverse momentum dependence of charged particle elliptic flow with $\sqrt{s_{NN}}=13.7, 17.3$ and $54.4$ GeV at different centrality bins in Au+Au collisions. Due to the errors of experimental data, an error 10\% is added to the predictions of the neural network as well. Since the predictions of $v_2$ given by the neural network model are smooth curves that vary with $p_T$, this model can be a promising approach for extrapolating elliptic flow behaviors in the $p_T$ range not covered by the experimental data.
	
	In Fig. \ref{pt_train}, the training results of the neural network for the transverse momentum distributions of charged particles with $\sqrt{s_{NN}}=7.7, 11.5, 14.5, 27, 39, 62.4$ and $200$ GeV at different centrality bins in Au+Au collisions are shown. It is observed that using the logarithmic transformation on the experimental data in the training set, the results of the neural network match up well with the experimental data. 
	
	In Fig. \ref{pt_test}, the test results of the neural network for the transverse momentum distributions of charged particles with $\sqrt{s_{NN}}=19.6$ and $130$ GeV at different centrality bins in Au+Au collisions are shown. As one can see, the neural network describes the experimental data well. This indicates that the trained neural network demonstrates low bias and low variance, allowing it to generate reliable predictions. It is also worth noting that the neural network captures the characteristic trend wherein particle yield at a given $p_T$ decreases as centrality increases, which indicates its capacity to model essential features of transverse momentum distributions of charged particles across varying collision parameters.
	
	In Fig. \ref{pt_prediction}, we show the predictions of the neural network for the transverse momentum distributions of charged particles with $\sqrt{s_{NN}}=9.2, 17.3$ and $54.4$ GeV at different centrality bins in Au+Au collisions. Because of the errors in experimental data, an error 10\% is added to the predictions of the neural network as well. These predictions indicate that by leveraging its training on transverse momentum distributions from a range of collision energies, the neural network is able to generalize and accurately predict particle distributions at RHIC energies, filling gaps in experimental data.
	
	Overall, it is observed that the neural network can be trained using three kinds of experimental data simultaneously, i.e. pseudorapidity density distributions, transverse momentum dependence of elliptic flow and transverse momentum spectra of charged particles, and achieve accurate training and test results, along with reasonable predictions across all observables.

	\section{Summary}\label{Sec4}
	
	In this work, we developed a physics-informed deep neural network to predict multiple bulk observables in Au+Au collisions at RHIC. The architecture is inspired by the space–time stages of a heavy-ion collision and incorporates locally connected hidden layers between QGP, chemical and kinetic freeze-out representations, together with a structured dual-input design encoding both initial collision characteristics and final-state kinematics. The network is trained exclusively on experimental measurements of charged-particle pseudorapidity density distributions, transverse-momentum spectra and elliptic flow over a broad range of collision energies and centralities. 
	
	We showed that this architecture simultaneously reproduces all three classes of observables with low training and test loss, and that both local connectivity and input splitting are essential for achieving this performance.
	Meanwhile, comparisons with alternative models show that our approach achieves superior performance, underscoring the importance of incorporating physical principles, such as local connections and structured input design, in the construction of effective neural network architectures. Using the trained model, we made predictions for charged particle pseudorapidity density distributions at $\sqrt{s_{NN}} = 17.3, 39$ and $54.4$ GeV,  charged particle elliptic flow at $\sqrt{s_{NN}} = 13.7, 17.3$ and $54.4$ GeV, charged particle transverse momentum distributions at $\sqrt{s_{NN}} = 9.2, 17.3$ and $54.4$ GeV in Au+Au collisions, energies not yet provided by the collaboration, and validated these predictions against a CLVisc hydrodynamic calculation with TRENTo initial conditions and through the global systematics of total charged-particle multiplicity per participant pair. 
	
	Our results demonstrate that physics-guided deep learning can act as an efficient, data-driven surrogate for multi-observable predictions in high-energy nuclear collisions, helping to fill gaps in heavy ion collisions and providing a flexible tool for phenomenological studies. It turns a fragmented database (from real experiments) into a continuous empirical surface in a space directly relevant for QGP phenomenology, and the mere fact that a single network (with proper physics-inspired structure) can reproduce multiple bulk observables across different collision energies and centralities in HICs suggests an empirical indication of underlying universality, which is compatible with the idea that the system is governed by a few number of effective physical macroscopic parameters (e.g., entropy/energy density, transport coefficients, freeze-out conditions, etc.), even though the network never sees those parameters explicitly.  Strictly speaking, our network is an empirical interpolator in the space of bulk observables and collision parameters: it reproduces regularities observed in the data, but does not constitute a microscopic theory of QCD matter. Future work will aim to extend this framework to additional observables (such as higher-order flow harmonics and identified-particle spectra), to other collision systems, and to incorporate systematic uncertainty estimates via Bayesian or ensemble approaches. Ultimately, such models may be integrated into global analyses as fast emulators for inferring QGP properties from multi-dimensional data sets.

	\begin{acknowledgments}
		We thank Prof. Wei Dai, Prof. Ze-Fang Jiang and Prof. Dirk H. Rischke for helpful discussions. This work is supported by the National Natural Science Foundation of China (NSFC) under Grant Nos. 12535010 and 92570117, the Ministry of Science and Technology of China under Grant No. 2024YFA1611004, the Shenzhen Peacock Fund under Grant No. 2023TC0179, the CUHK-Shenzhen University Development Fund under Grant Nos. UDF01003041 and UDF03003041, the China Scholarship Council (CSC) under Grant No. 202406770003, and the China Postdoctoral Science Foundation under Grant No. 2025M781519.
	\end{acknowledgments}

	\appendix
	\section{CLVisc model with T\raisebox{-0.5ex}{R}ENTo initial condition}\label{appendixA}
	
	 In the following, we briefly introduce the CLVisc model with T\raisebox{-0.5ex}{R}ENTo. For more details, we refer to Refs. \cite{Pang2018,Wu2022,Moreland2015}.
	 
	\subsection{Initial condition}
\begin{figure}
		\includegraphics[scale=0.7]{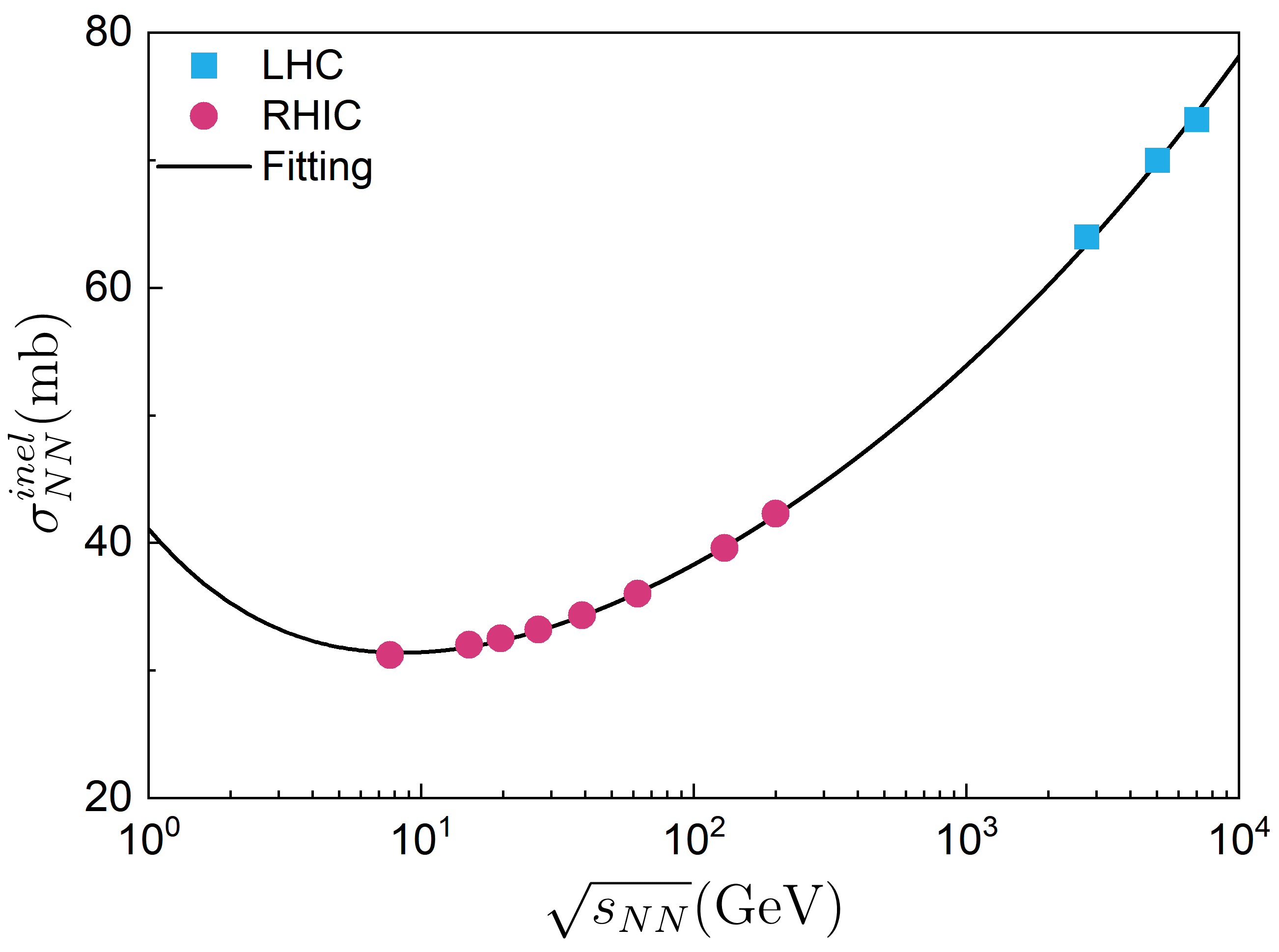}
		\caption{\label{cross_section}The NN inelastic collision cross section versus $\sqrt{s_{NN}}$. The circles and squares are the experimental results from RHIC \cite{Adare2016} and LHC \cite{Aad2012,Abelev2013,Abelev2013a}, respectively. The line is the fitting result from Eq. \eqref{eq3}.}
	\end{figure}

	 T\raisebox{-0.5ex}{R}ENTo is a model for the initial conditions of high-energy nuclear collisions and can generate realistic Monte Carlo initial entropy profiles without assuming specific physical mechanisms for entropy production, pre-equilibrium dynamics or thermalization. However, T\raisebox{-0.5ex}{R}ENTo requires the nucleon-nucleon collision inelastic cross-section. In this work, we adopt the parameterization \cite{Donnachie1992}, 
	\begin{equation}\label{eq3}
		\begin{aligned}
			\sigma^{inel}_{NN} = 17.2031s_{NN}^{0.1641}+23.8981s_{NN}^{-0.5783},
		\end{aligned}
	\end{equation}
	where $\sqrt{s_{NN}}$ is the collision energy in units of GeV. This parameterization is shown in Fig. \ref{cross_section}, which demonstrates that it can fit the experimental data well. With this fitting result, T\raisebox{-0.5ex}{R}ENTo can be used to produce the number of participants at the collision energy that we need. 	The inelastic cross section used in this paper is already listed in Tab. \ref{t3}.
	
	In order to describe the charged particle pseudorapidity density distributions produced in the most central collision, the initial transverse entropy profile is assumed to be the following form \cite{Zhu2025},
	\begin{equation}
		\begin{aligned}
			\left.\frac{d S}{d y}\right|_{\tau=\tau_{0}} = K \cdot T_{R},
		\end{aligned}
	\end{equation}
	where $\tau_0$ is the start proper time for hydrodynamic evolution, $K$ is the scale factor and $T_R$ is the generalized average of the thickness functions of two nuclei \cite{Moreland2015}. To describe the plateau structure in the longitudinal direction, the following function is used \cite{Zhu2025},
	\begin{equation}
		\begin{aligned}
			H(\eta)=\exp \left[-\frac{\left(\eta-\eta_{flat}\right)^{2}}{2 \eta_{gw}{ }^{2}} \theta\left(|\eta|-\eta_{flat}\right)\right],
		\end{aligned}
	\end{equation}
	where $\eta_{flat}$ and $\eta_{gw}$ are parameters. The parameters used for the initial conditions, including $\tau_0$, $K$, $\eta_{flat}$, and $\eta_{gw}$, are tuned to match the charged particle pseudorapidity density distributions in the most central collisions. The specific values are listed in Tab. \ref{t3}.
	
	\begin{table}[!htb]
		\renewcommand{\arraystretch}{1.5}
		\centering
		\caption{The parameters of the initial condition used for Au+Au collisions at different collision energies.}
		\begin{tabular}{>{\centering\arraybackslash}p{1.9cm}
				>{\centering\arraybackslash}p{1.7cm}
				>{\centering\arraybackslash}p{1.7cm}
				>{\centering\arraybackslash}p{1.7cm}}
			\hline\hline
			\textbf{Parameters} & \textbf{54.4 GeV} & \textbf{39 GeV} & \textbf{17.3 GeV} \\
			\hline
			$\sigma^{\mathrm{inel}}_{NN}\,(\mathrm{fm}^2)$ & 3.55 & 3.43 & 3.21 \\
			$\tau_0\,(\mathrm{fm})$ & 0.6 & 0.6 & 0.6 \\
			$K$ & 45 & 42 & 40 \\
			$\eta_{\mathrm{flat}}$ & 2.5 & 2.2 & 1.8 \\
			$\eta_{\mathrm{gw}}$ & 0.1 & 0.07 & 0.04 \\
			\hline\hline
		\end{tabular}
		\label{t3}
	\end{table}
	
	\subsection{Hydrodynamic evolution}
	
	After obtaining the initial condition generated from the T\raisebox{-0.5ex}{R}ENTo, the CLVisc model is used to simulate the hydrodynamic evolution of the collision system through the following energy momentum conservation equation,
	\begin{align}
		\nabla_\mu T^{\mu\nu} &= 0,
	\end{align}
	where $T^{\mu\nu}$ is the energy-momentum tensor which is defined as follows,
	\begin{align}
		T^{\mu\nu} &= \varepsilon u^\mu u^\nu - p\Delta^{\mu\nu} + \pi^{\mu\nu},
	\end{align}
	where $\varepsilon$ is the energy density, $u^\mu$ is the fluid four-velocity, $p$ is the pressure, $\Delta^{\mu\nu} = g^{\mu\nu} - u^\mu u^\nu$ is the projection operator and $\pi^{\mu\nu}$ is the shear viscosity tensor. $\pi^{\mu\nu}$ is defined based on an Israel-Stewart-like second-order hydrodynamic expansion \cite{Jiang2023},
	\begin{equation}
		\begin{aligned}
			\Delta_{\alpha \beta}^{\mu \nu} u^\sigma \partial_\sigma \pi^{\alpha \beta} = & -\frac{1}{\tau_{\pi}}\left(\pi^{\mu \nu} - \eta_{v} \sigma^{\mu \nu}\right)-\frac{4}{3} \pi^{\mu \nu} \theta \\ & - \frac{5}{7} \pi^{\alpha<\mu} \sigma_{\alpha}^{\nu>} + \frac{9}{70} \frac{4}{e+P} \pi_{\alpha}^{<\mu} \pi^{\nu>\alpha},
		\end{aligned}
	\end{equation}
	where $\Delta_{\alpha \beta}^{\mu \nu}=\frac{1}{2}\left(\Delta_{\alpha}^{\mu} \Delta_{\beta}^{\nu}+\Delta_{\alpha}^{\nu} \Delta^{\mu}{ }_{\beta}\right)-\frac{1}{3} \Delta^{\mu \nu} \Delta_{\alpha \beta}$ is  second-order symmetric projection operator, $\tau_{\pi}$ is the relaxation times, $\eta_v$ is the shear viscosity coefficient, $\theta = \partial_\mu u^\mu$ is the expansion rate and $\pi^{<\mu\nu>} = \Delta_{\alpha \beta}^{\mu \nu} \pi^{\alpha\beta}$ is the traceless symmetric tensor. The relaxation times, $\tau_{\pi}$, and shear viscosity coefficient, $\eta_v$, have the following relation with the specific shear viscosity, $C_{\eta_v}$, that is treated as a model parameter \cite{Jiang2023},
	\begin{align}
		\tau_{\pi}=\frac{5 C_{\eta_{v}}}{T},\\
		C_{\eta_{\mathrm{v}}}=\frac{\eta_{v} T}{e+P}.
	\end{align}
	The specific shear viscosity, $C_{\eta_v}$, adopted in this paper is 0.08.
	
	In this work, the bulk pressure and baryon current were not considered. The equation of state used is from the HotQCD Collaboration \cite{Bazavov2014}. When the energy density of the evolving system is lower than the freeze-out energy density ($e_{frz}=0.4$ ${\rm GeV/fm^3}$) \cite{Jiang2023}, the Cooper–Frye formalism \cite{Jiang2023} is used to obtain the hadron momentum distribution. The resonance decays have been considered. The centrality classes are divided based on the total entropy generated by T\raisebox{-0.5ex}{R}ENTo.
	
	\bibliography{Ref_v13.bib}
	
\end{document}